\documentclass[a4paper,11pt]{article}
\PassOptionsToPackage{dvipdfmx}{graphicx}
\usepackage{pos}
\usepackage{doi}
\usepackage{hyperref}

\title{Large $N$ simulation of the twisted reduced matrix model with an adjoint Majorana fermion}
\ShortTitle{Large $N$ simulation of the twisted reduced matrix model ...}

\author[a,b]{Pietro~Butti}
\author[a]{Margarita~Garc\'{i}a~P\'{e}rez}
\author[a,b]{Antonio~Gonz\'{a}lez-Arroyo}
\author*[c,d]{Ken-Ichi~Ishikawa}
\author[d]{Masanori~Okawa}

\affiliation[a]{Instituto de F\'{i}sica Te\'{o}rica UAM-CSIC, Nicol\'{a}s Cabrera 13-15, Universidad Aut\'{o}noma de Madrid, Cantoblanco, E-28049, Madrid, Spain}

\affiliation[b]{Departamento de F\'{i}sica Te\'{o}rica, M\'{o}dulo 15, Universidad Aut\'{o}noma de Madrid, Cantoblanco, E-28049, Madrid, Spain}
\affiliation[c]{Core of Research for the Energetic Universe, Graduate School of Advanced Science and Engineering, Hiroshima University, Higashi-Hiroshima, Hiroshima 739-8526, Japan}
\affiliation[d]{Graduate School of Advanced Science and Engineering, Hiroshima University, Higashi-Hiroshima, Hiroshima 739-8526, Japan}

\emailAdd{ishikawa@theo.phys.sci.hiroshima-u.ac.jp}

\abstract{
To investigate the properties of the large $N$ limit of $\mathcal{N} = 1$ SUSY Yang-Mills theory, we have started a
study for a reduced matrix model with an adjoint Majorana fermion. The gauge action is based
on the Wilson action and the adjoint-fermion one is the Wilson-Dirac action on a reduced lattice with twisted
gauge boundary condition. We employ the RHMC algorithm in which the absolute value of the Pfaffian is
incorporated. The sign of the Pfaffian is included with the re-weighting method and separately measured as
an observable. In this talk, we show the configuration generation status towards the large $N$ limit and 
the behavior of the lowest/lower eigenvalue(s) of the Wilson-Dirac
adjoint fermion operator. We investigated the sign of the Pfaffian and the critical hopping parameters for the 
chiral limit. The sign of the Pfaffian is always positive on the configurations we have generated.
The critical hopping parameters derived from the eigenvalues of the Dirac operator are consistent with those
derived from the PCAC mass relation with non-singlet flavor adjoint fermions.
}

\FullConference{%
 The 38th International Symposium on Lattice Field Theory, LATTICE2021
  26th-30th July, 2021
  Zoom/Gather@Massachusetts Institute of Technology
}

\begin{document}

\begin{flushright}
    \vspace*{-8em}
FTUAM-21-5, HUPD-2110, IFT-UAM/CSIC-21-121
    \vspace*{4em}
\end{flushright}

\maketitle

\section{Introduction}

Understanding the non-perturbative dynamics of Yang-Mills theories is one of the important open problems in theoretical physics. 
One of the strategies to investigate this issue in a simplified set up is the $1/N$ expansion. In the limit of large number of colours, 
Yang-Mills theory retains many of the non-perturbative aspects of finite $N$ theories while exhibiting a number of simplified features, as planarity or volume 
independence. In this work, we combine the large $N$ limit with  lattice techniques to investigate non-perturbatively 
${\cal N}=1$ SUSY SU($N$) Yang-Mills theory.

The approach we take for this study is based on the property known as volume reduction.
Lattice studies of pure SU($N$) Yang-Mills indicate that finite volume effects disappear in the large $N$ limit, 
allowing to simulate the theory on a one-point lattice endowed with twisted boundary conditions~\cite{Eguchi:1982nm,GonzalezArroyo:1982ub,GonzalezArroyo:1982hz}. 
Reduction can as well be extended to include dynamical fermions in the adjoint representation~\cite{Gonzalez-Arroyo:2013gpa}, in particular 
in the case of a single massless Majorana fermion that corresponds to ${\cal N}=1$ SUSY.
Although the lattice breaks supersymmetry, 
${\cal N}=1$ SUSY gets restored in the chiral and continuum limits~\cite{Curci:1986sm}.
This restoration can be investigated by analyzing the mass spectrum and the SUSY Ward--Takahashi identity. 
In this work we followed an alternative approach based on the analysis of the eigenspectrum of the adjoint Wilson fermion operator
and the PCAC mass of the non-singlet flavor pseudo-scalar--axial current channel.
Here we present an account of the methodology employed and the status
of the simulations; a preliminary study of scale setting for our results 
has been presented at this conference by P. Butti~\cite{Pietro}.

The paper is organized as follows. Next section is devoted to explain
the model and the methodology.
In Section~\ref{sec:2}, we show the simulation results including
the complex eigenvalues of the adjoint Wilson fermion operator,
the minimum absolute value of the eigenvalue of the adjoint Hermitian Wilson fermion operator,
and the chiral limit of the eigenvalue and the PCAC mass.
We summarize our results in the last section.

\section{Model and Simulation method}
\subsection{Model}
Our model is a volume reduced version of the lattice model with Wilson gauge and one adjoint Majorana Wilson
fermion actions with twisted boundary conditions.
The partition function and the action are defined by
\begin{align}
  {\cal Z} & =\int \prod_{\mu=1}^{4} dU_{\mu} \mathrm{Pf}[CD_W]e^{-S_G[U]},
             \label{eq:partitionfunction}
  \\
  S_G[U] &=  b N \sum_{\mu,\nu=1,\mu\ne\nu}^{4} \mathrm{Tr}\left[ I - z_{\mu\nu} U_{\mu}U_{\nu}U_{\mu}^{\dag}U_{\nu}^{\dag}\right],
             \label{eq:TEKaction}
  \\
  D_W&= I - \kappa_{\mathrm{adj}}\sum_{\mu=1}^{4}\left[ (1-\gamma_{\mu}) V_{\mu} + (1+\gamma_\mu)V_{\mu}^{T}\right],
             \label{eq:AdjWDmatrix}
\end{align}
where $U_\mu$ and  $V_\mu$ are SU($N$) matrices in the fundamental representation and in the adjoint representation, respectively.
The gauge action \eqref{eq:TEKaction} is the so-called twisted-Eguchi-Kawai (TEK)
action~\cite{Eguchi:1982nm,GonzalezArroyo:1982ub,GonzalezArroyo:1982hz}.
Twisted gauge boundary conditions are imposed through the twist phase factor $z_{\mu\nu}$ defined by
\begin{align}
 z_{\mu\nu} =\exp\left[\dfrac{2\pi k i}{\sqrt{N}}\right], \quad z_{\nu\mu}=z_{\mu\nu}^{*} \qquad \mbox{for $\mu < \nu$},
\end{align}
where $k$ is an integer coprime to $\sqrt{N}$.
$D_W$ is the Wilson-Dirac fermion matrix in the adjoint representation, $C=\gamma_4\gamma_2$ is the charge conjugation gamma matrix, and
$\mathrm{Pf}[X]$ is the Pfaffian for a skew-symmetric matrix $X^T=-X$.
Using the properties that $V_{\mu}^{\dag}=V_{\mu}^{T}$, $C^{T}=-C$, and  $C\gamma_\mu C^{T}=-\gamma_\mu^T$, we can prove
the skew-symmetry as $(CD_W)^T=-CD_W$. The weight $\mathrm{Pf}[CD_W]$ in the path integral given by eq.~\eqref{eq:partitionfunction}
introduces one dynamical adjoint Majorana fermion in the system.

We employ a Markov chain Monte Carlo (MCMC) algorithm to generate configurations from \eqref{eq:partitionfunction}.
To do this, we need to include the Pfaffian in the MCMC algorithm.
In general it is proved that the Pfaffian is real but  positivity is not guaranteed.
In order to avoid a negative sign probability weight for MCMC, we transform the partition function as follows:
\begin{align}
  {\cal Z}
  &=\int \prod_{\mu=1}^{4} dU_{\mu}\mathrm{sign}\left(\mathrm{Pf}[CD_W]\right) \left|\mathrm{det}[Q_W^2]\right|^{1/4}e^{-S_G[U]},
             \label{eq:partitionfunction2}
\end{align}
where $Q_W = D_W\gamma_5$ and the property $\left|\mathrm{Pf}[CD_W]\right|=\left|\mathrm{det}[Q_W^2]\right|^{1/4}$ is used.
We can then apply MCMC algorithms to \eqref{eq:partitionfunction2} with a positive weight
$\left|\mathrm{det}[Q_W^2]\right|^{1/4}e^{-S_G[U]}$. The sign of the Pfaffian should be incorporated in measured observables
using re-weighting.

\subsection{Simulation method}
We employ the rational Hybrid Monte Carlo (RHMC) algorithm~\cite{Kennedy:1998cu,Clark:2003na} to
incorporate the fractional power of the fermionic determinant. 
To attain that goal $\left|\mathrm{det}[Q_W^2]\right|^{1/4}$  is rewritten in the following pseudo-fermionic integral form:
\begin{align}
  \left|\mathrm{det}[Q_W^2]\right|^{1/4}& =\int d\phi d\phi^{\dag} e^{-S_Q}\quad \mbox{with} \quad
  S_Q = \mathrm{Tr}\left[\phi^{\dag} R_{N_R}^{(-1/4)}(Q_W^2) \phi\right],
\end{align}
where $\phi$ is the pseudo-fermion field in bi-fundamental form for the adjoint representation.
When we apply the Hermitian Wilson-Dirac matrix $Q_W$ to the pseudo-fermion field $\phi$ in bi-fundamental form,
we get 
\begin{align}
  Q_W \phi = D_W \gamma_5 \phi&= \gamma_5 \phi -
             \kappa_{\mathrm{adj}}\sum_{\mu=1}^{4}\left[ (1-\gamma_{\mu})\gamma_5 U_{\mu}\phi U_{\mu}^{\dag}
             + (1+\gamma_\mu)\gamma_5 U_{\mu}^{\dag}\phi U_{\mu}\right],
  \label{eq:AdjQWbifmatrix}
\end{align}
where the traceless condition $\mathrm{Tr}\phi = 0$ is imposed to remove the unwanted U(1) contribution  before the multiplication.
$R_{N_R}^{(p)}(x)$ is the $N_R$-th order rational polynomial approximation to $x^{p}$ defined by 
\begin{align}
  x^{p} \simeq R_{N_R}^{(p)}(x) \equiv \alpha^{(p)}_0 + \sum_{j=1}^{N_R}\dfrac{\alpha_j^{(p)}}{x-\beta_j^{(p)}}.
  \label{eq:PADE}
\end{align}
The rational approximation \eqref{eq:PADE} for $(Q_W^2)^p\phi$ requires the inversion of
$(Q_W^2 - \beta^{(p)}_j)^{-1}\phi$ for which the multi-shift conjugate gradient algorithm is employed.

The parameters $\{\alpha_j^{(p)},\beta_j^{(p)}\}$ are determined to minimize the metric:
$\max_{x \in [a,b]}\left|(x)^p - R_{N_R}^{(p)}(x)\right|$, for a real number interval $x\in[a,b]$.
To minimize the metric for the matrix $Q_W^2$, the lowest and largest eigenvalues of $Q_W^2$ should be monitored.
In the RHMC algorithm we monitor the exterior eigenvalues at the beginning and the end of each
molecular dynamics (MD) evolution and choose the polynomial order $N_R$ to satisfy the metric below double precision tolerance $10^{-15}$.
With the parameters, the value of the total Hamiltonian is precisely evaluated for the Metropolis test.
To compute the exterior eigenvalues of $Q_W^2$,
which has the symmetry $B(Q_W^2)B^{-1}=(Q_W^2)^T$ with $B=\gamma_5 C$ resulting in a two-folded eigenspectrum,
we employ an improved thick-restart Lanczos type algorithm~\cite{Ishikawa:2020xac} to treat the degeneracy.

On the other hand, the rational polynomial parameters $\{\alpha_j^{(p)},\beta_j^{(p)}\}$ and
the order $N_R$ have to be the same for every MD evolution trajectory to make the RHMC algorithm exact.
The polynomial order for the MD evolution and eigenvalue interval $[a,b]$,
and then the parameters $\{\alpha_j^{(p)},\beta_j^{(p)}\}$,
are tuned during the thermalization steps to sufficiently cover the measured exterior eigenvalues
with a low approximation tolerance,
and the tuned value is kept for all MD trajectories in the production run.

The sign of the Pfaffian can be determined by counting the number of negative real eigenvalues
of $D_W$~\cite{Demmouche:2010sf,Bergner:2011zp,Bergner:2011wf}.
On each configuration $U_\mu$ generated by the RHMC algorithm,
we compute the complex eigenvalues of $D_W$ near the origin of the complex plane 
using ARPACK~\cite{ARPACKURL,lehoucq1998arpack} to determine
the sign used in the reweighting method when evaluating observables.

\begin{table}[t]
    \centering
    \begin{tabular}{cccc} \hline
      $b$  & $(N,k)$                   & $\kappa_{\mathrm{adj}}$                  \\\hline
     0.360 & $(289,5)$                 & $0.1760, 0.1780, 0.1800, 0.1820, 0.1840$ \\
     0.350 & $(169,5),(289,5),(361,7)$ & $0.1775, 0.1800, 0.1825, 0.1850, 0.1875$ \\
     0.340 & $(169,5),(289,5),(361,7)$ & $0.1850, 0.1875, 0.1890, 0.1910, 0.1930$ \\ \hline
    \end{tabular}
    \caption{Model parameters.
       We accumulated over 600 configurations for each pair of $(N,k)$ and $\kappa_{\mathrm{adj}}$.}
    \label{tab:params}
\end{table}

\section{Simulation results}
\label{sec:2}
\subsection{Model Parameters}  
The TEK model exhibits a first order phase transition separating the weak and strong coupling regions.
To approach the proper continuum limit, simulations at finite lattice cut-off should be performed in the weak coupling region.
For the current model \eqref{eq:partitionfunction}, with a single adjoint Majorana fermion,
a first order phase transition line separating the weak and strong coupling regions remains.
After a brief survey of the parameter space $\{b,\kappa_{\mathrm{adj}}\}$, we roughly determined the position of this line
and have generated configurations with parameters pertaining to the weak coupling region as tabulated in Table~\ref{tab:params}.
Details on the scale setting and the meson spectrum with fundamental fermions
have been presented at this conference by P.~Butti and can be found in~\cite{Pietro}.

We employ the RHMC algorithm described in the previous section for the configuration generation.
Each configuration is separated with five trajectories with a trajectory length of $\tau=1$
and with HMC acceptance rates greater than 70\%.
Numerical computations were done on the following computer systems:
(i) SX-ACE at Osaka University,
(ii) Oakbridge-CX at University of Tokyo, and
(iii) Subsystem B of ITO system at Kyushu University.
The code for the RHMC algorithm is primarily written in Fortran 2003.

\begin{figure}[t]
    \centering
    \includegraphics[clip,scale=0.045]{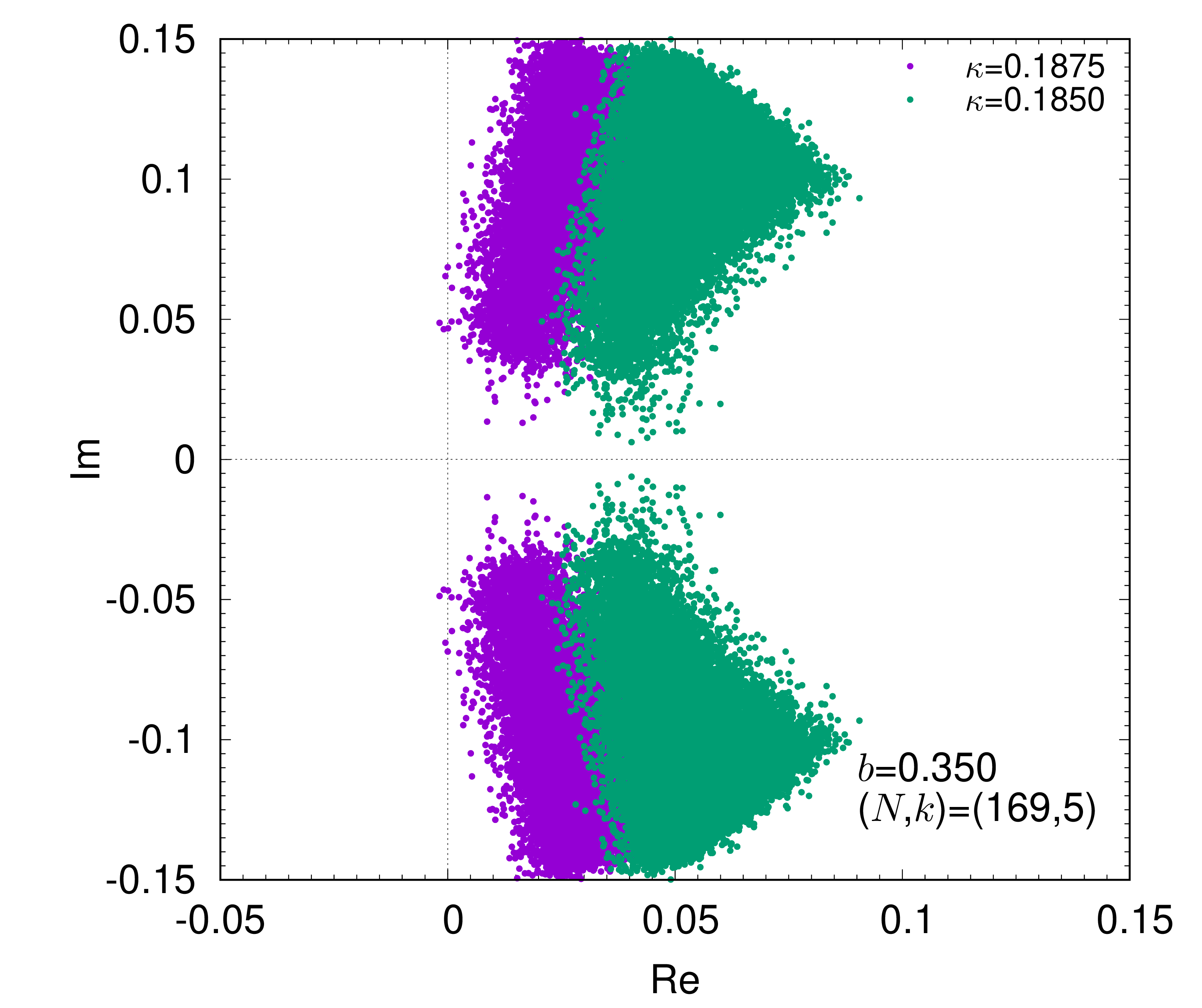}
    \includegraphics[clip,scale=0.045]{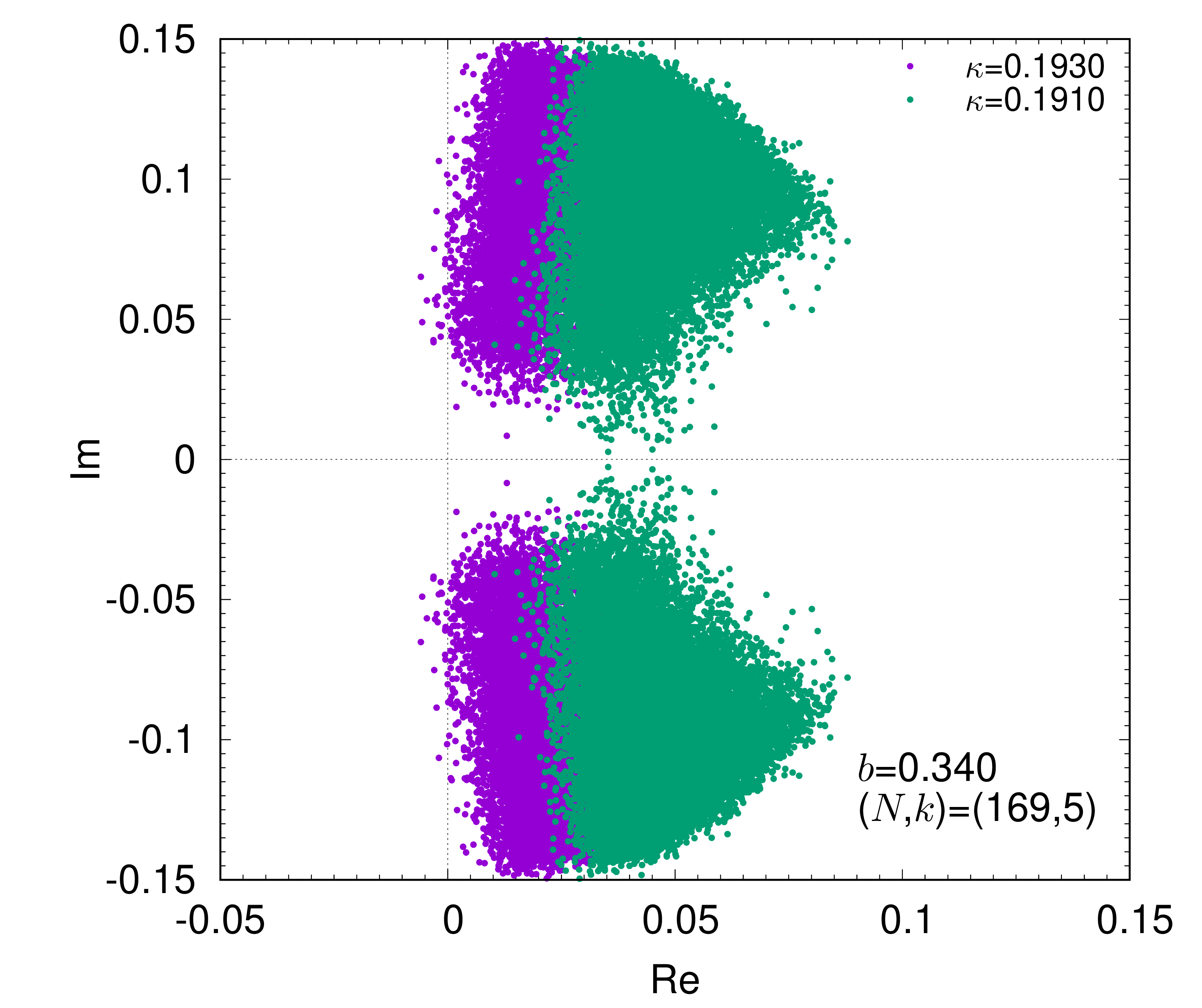}\\
    \includegraphics[clip,scale=0.045]{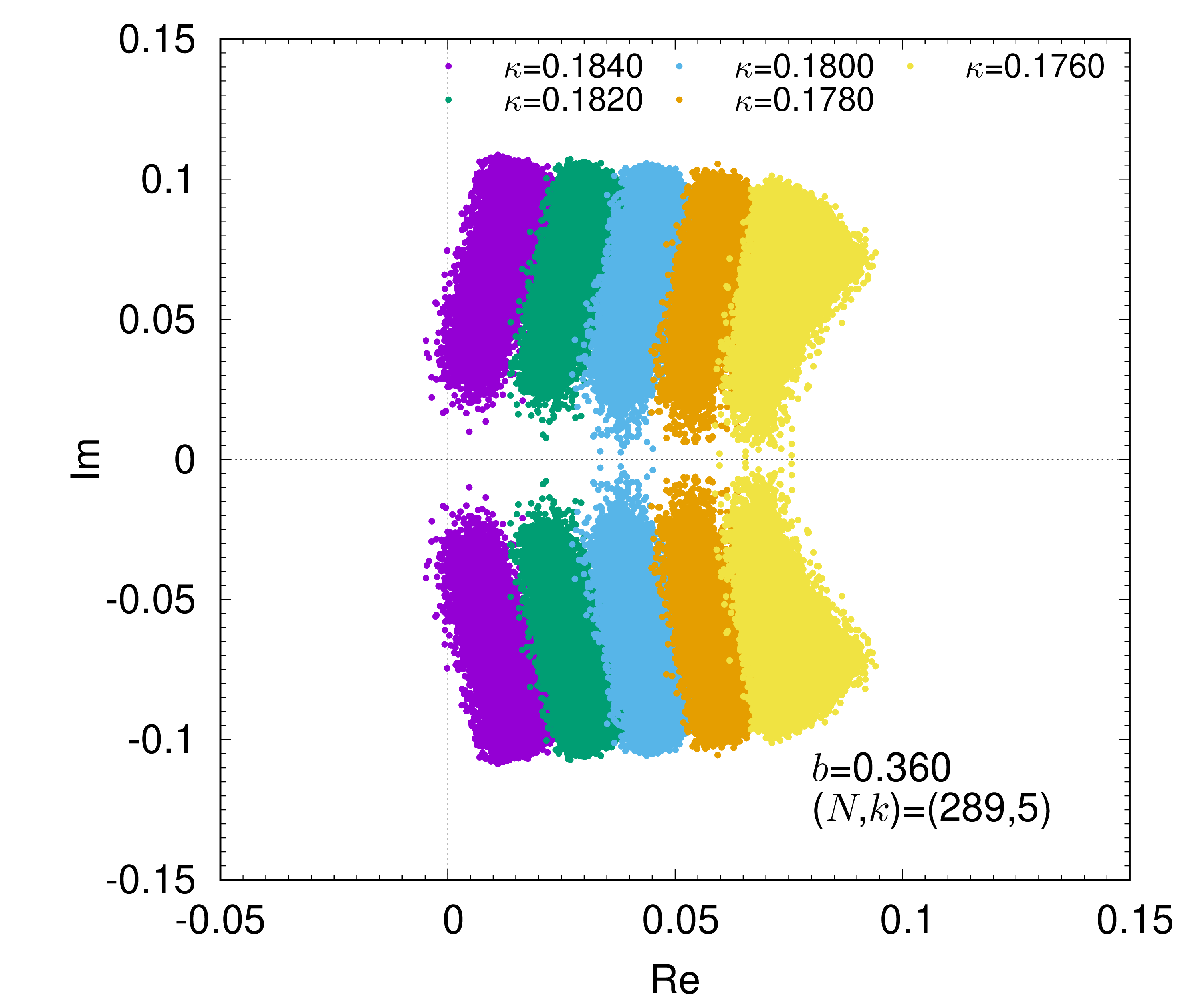}
    \includegraphics[clip,scale=0.045]{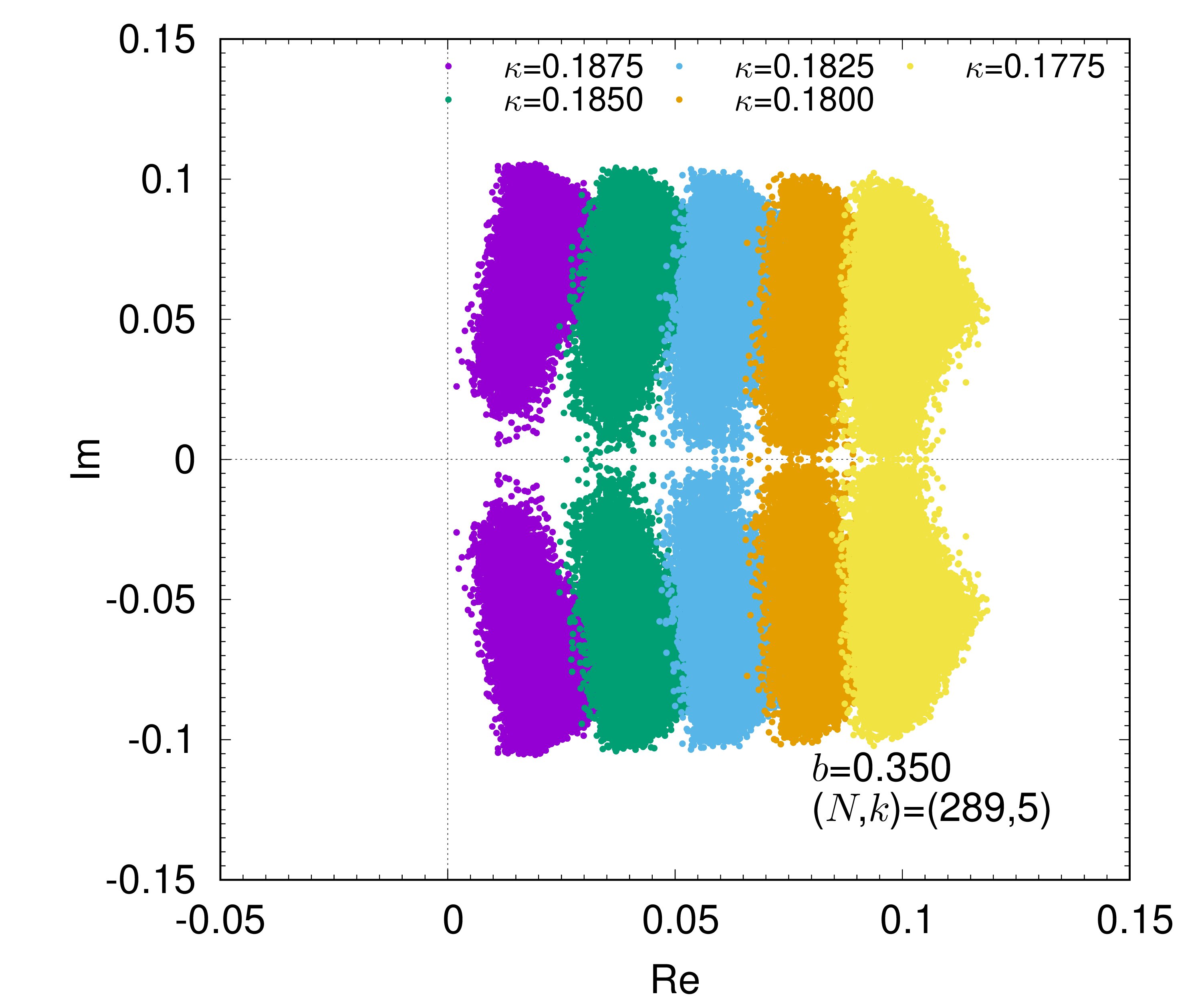}
    \includegraphics[clip,scale=0.045]{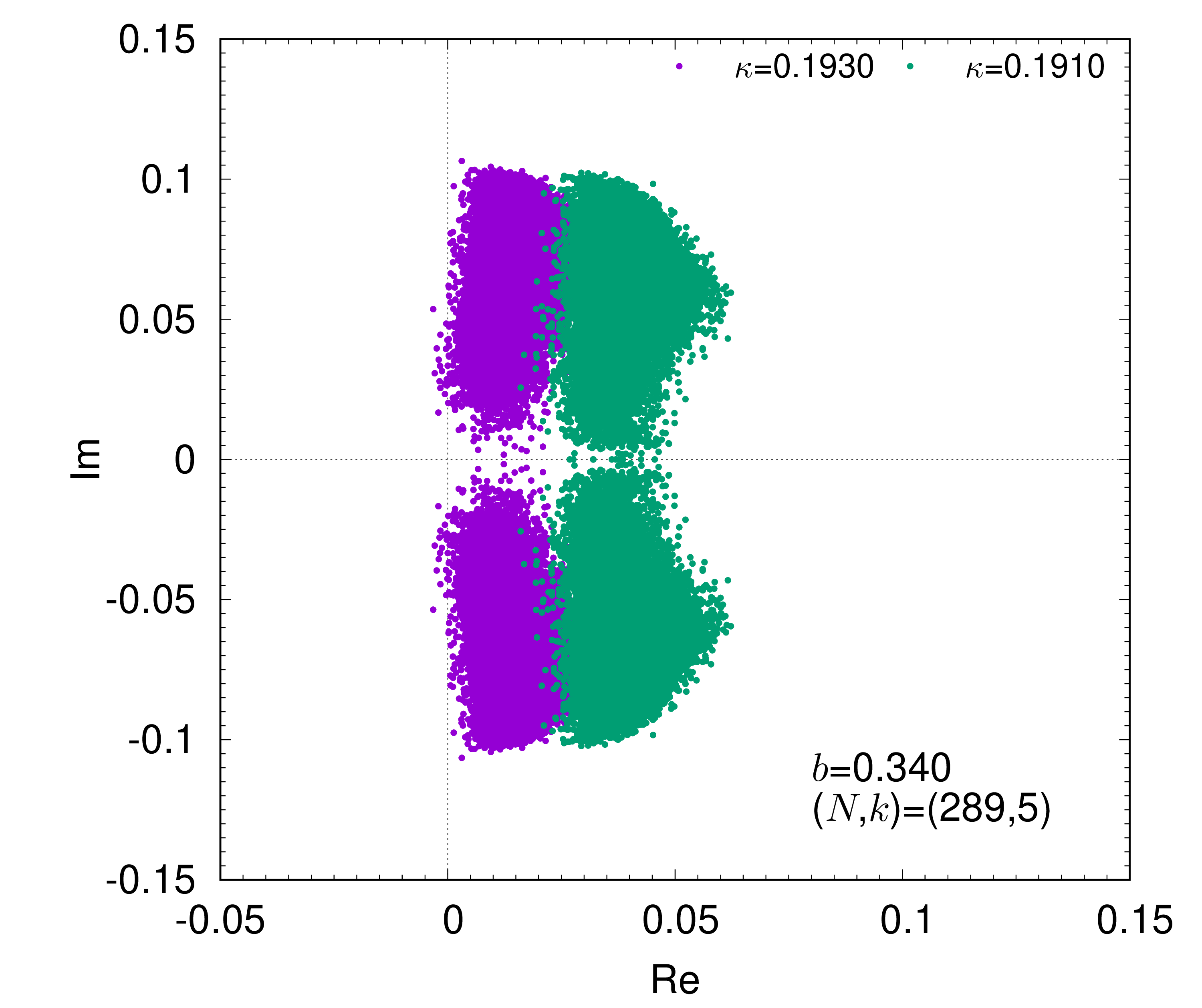}\\
    \includegraphics[clip,scale=0.045]{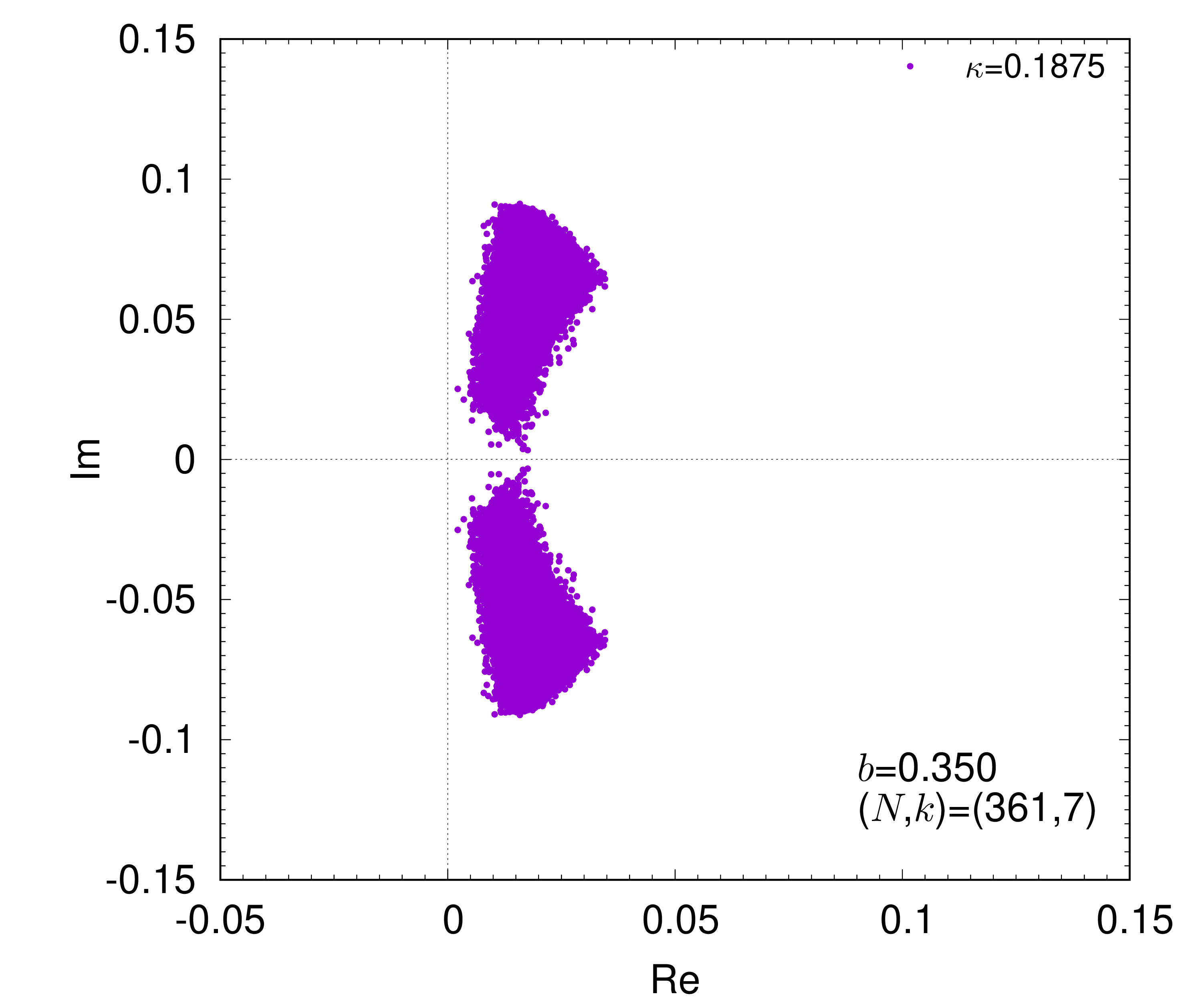}
    \includegraphics[clip,scale=0.045]{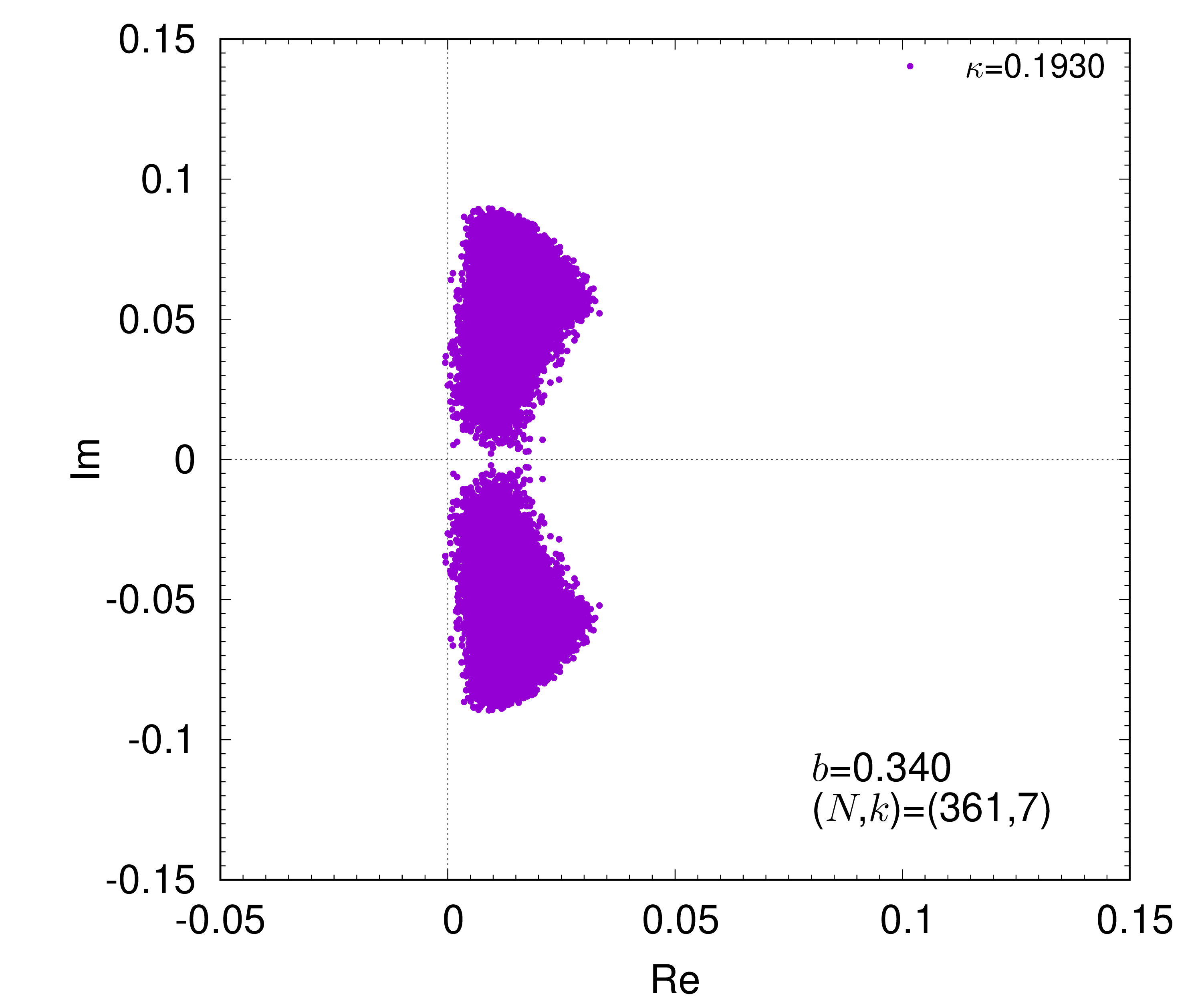}
    \caption{Complex eigenvalues of $D_W$ (Top: $N=169$, Middle: $N=289$, Bottom: $N=361$).}
    \label{fig:complexeigen}
\end{figure}

\subsection{Sign of the Pfaffian, lowest eigenvalues, and chiral limit}
We evaluated the sign of the Pfaffian by counting the number of negative real eigenvalues of $D_W$.
Figure~\ref{fig:complexeigen} shows the eigenvalue distribution of $D_W$ in the complex plane.
We compute 100 eigenvalues close to $z=-0.1$ in the complex plane with the shift-invert mode of ARPACK on each configuration,
and all eigenvalues are overlaid in the plane.
At the lightest adjoint fermion mass at each $b$ and $(N,k)$, we did not observe any negative real eigenvalues.
As expected, for heavier fermion masses, negative real eigenvalues are
also not present, as we explicitly checked for various cases.
We conclude that for our simulation parameters in Table~\ref{tab:params}, the sign of the Pfaffian is always positive.

\begin{figure}[t]
    \centering
    \includegraphics[clip,scale=0.49]{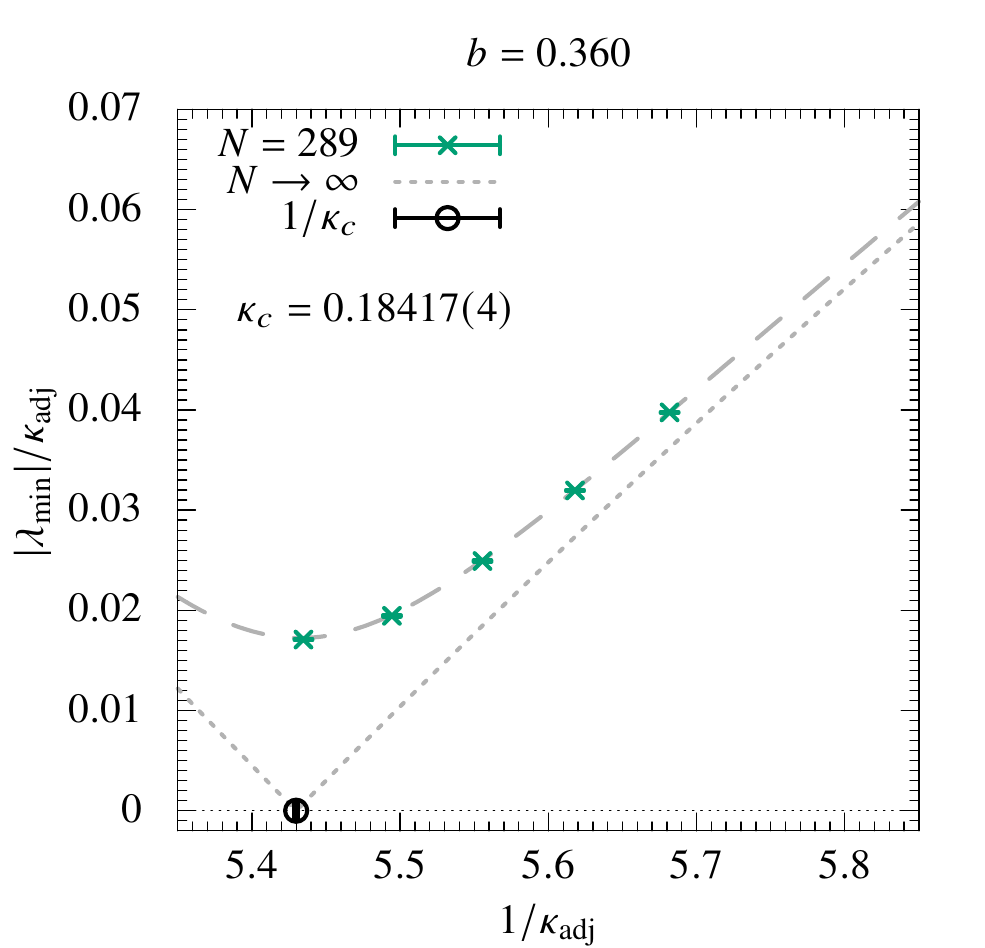}
    \includegraphics[clip,scale=0.49]{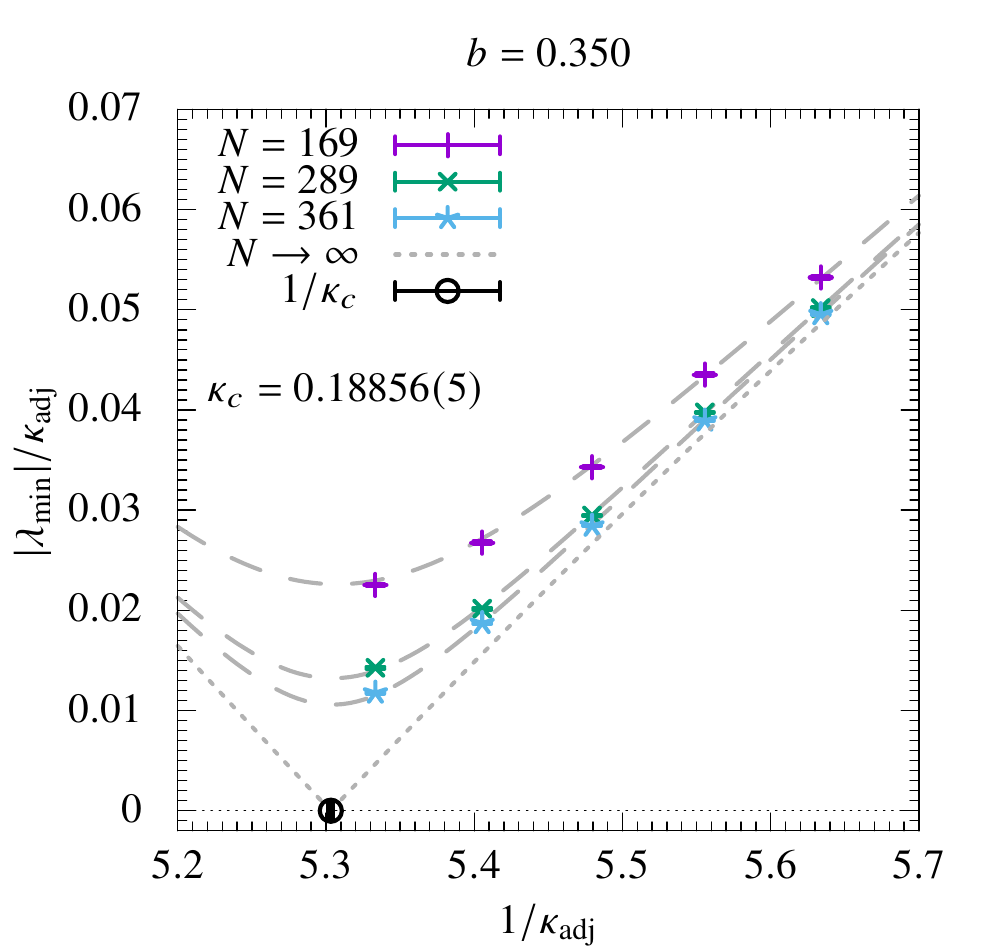}
    \includegraphics[clip,scale=0.49]{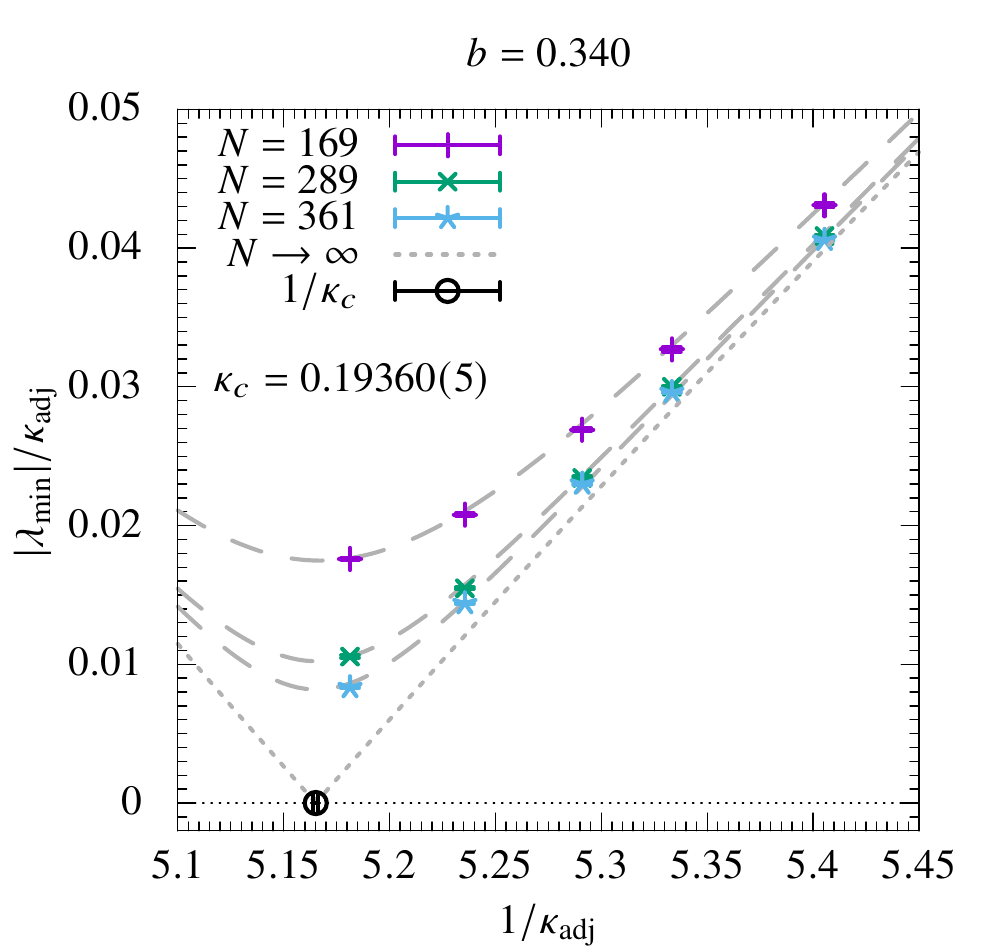}
    \caption{$\kappa_{\mathrm{adj}}$ dependence of $|\lambda_{\mathrm{min}}|$, with $\lambda_{\mathrm{min}}$ the eigenvalue 
of $Q_W$ with lowest absolute value.}
    \label{fig:lambdavskappa}
\end{figure}

The dependence on $\kappa_{\mathrm{adj}}$ of the expectation value of $|\lambda_{\mathrm{min}}|$, corresponding to the eigenvalue 
of $Q_W$ with lowest absolute value, is displayed on Figure~\ref{fig:lambdavskappa}.
We expect that $|\lambda_{\mathrm{min}}|$ is a linear function of the mass of the Majorana fermion, with vanishing value
corresponding to the chiral limit.
The $\kappa_{\mathrm{adj}}$ dependence is fitted with the following functional form:
\begin{align}
  \dfrac{\left|\lambda_{\mathrm{min}}\right|^2}{\left(\kappa_{\mathrm{adj}}\right)^2}
  = A\left(\dfrac{1}{\kappa_{\mathrm{adj}}}-\dfrac{1}{\kappa_{c}}\right)^2 + \dfrac{\delta}{N^2},
  \label{eq:fitfunc}
\end{align}
where, as explained earlier,  $\left|\lambda_{\mathrm{min}}\right|^2$ is the expectation value of the lowest eigenvalue of $Q_W^2$
measured during the RHMC algorithm.
The last term in the right hand side of \eqref{eq:fitfunc} represents the finite $N$ (or finite volume correction
with twisted boundary conditions) and $\kappa_c$ is the critical hopping parameter.
$A, \kappa_{c}$ and $\delta$ are fitting parameters.
We fit all the data simultaneously as a function of $(\kappa_{\mathrm{adj}}, N)$ at each $b$, and
this functional form describes well all our data.

\begin{figure}[t]
    \centering
    \includegraphics[clip,scale=0.80]{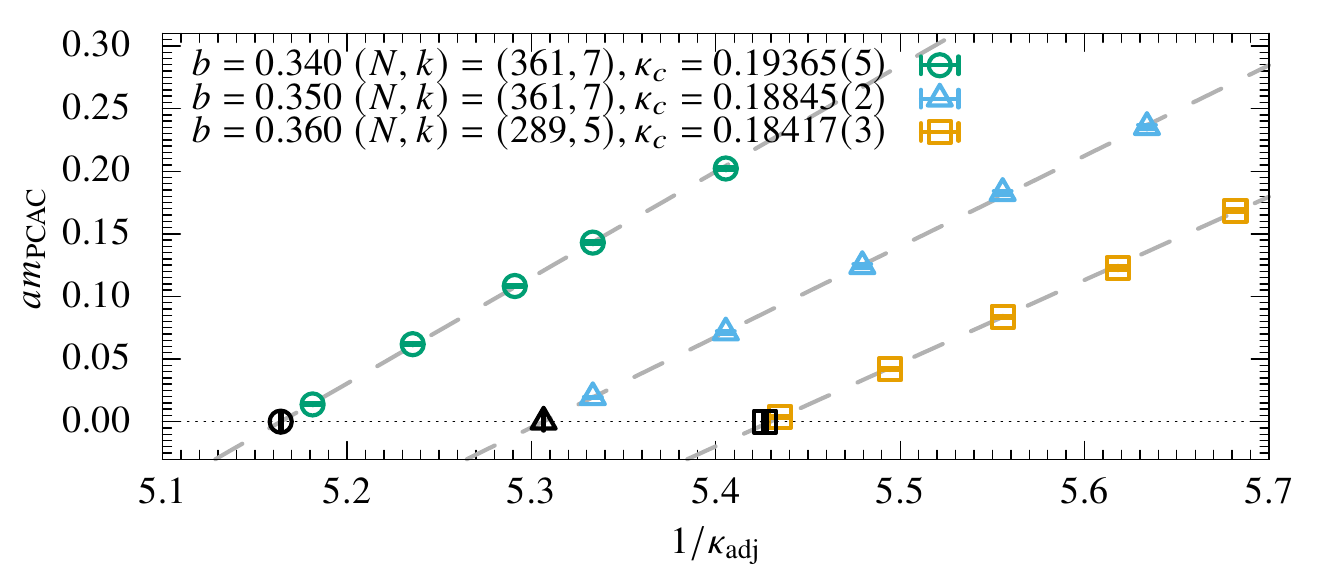}
    \caption{$\kappa_{\mathrm{adj}}$ dependence of the PCAC mass.}
    \label{fig:PCAC}
\end{figure}

The critical hopping parameters from $|\lambda_\mathrm{min}|$  can be compared
with those determined from the PCAC mass of adjoint Dirac-fermion meson correlation functions~\cite{Ali:2019agk}.
We construct non-singlet flavor meson correlation functions made of adjoint Wilson-Dirac fermions~\cite{Gonzalez-Arroyo:2015bya,Perez:2020vbn}.
The $\kappa_{\mathrm{adj}}$ dependence of the PCAC mass from the pseudo-scalar--axial vector channel is shown in Figure~\ref{fig:PCAC},
where the results at the largest value of $N$ are extrapolated linearly in $1/\kappa_{\mathrm{adj}}$ for each value of $b$.
The resulting critical hopping parameters are given by:
\begin{alignat}{3}
    \kappa_c &= 0.19365(5)&\quad& \mbox{[PCAC]},&\quad  0.19360(5) \quad\mbox{[|$\lambda_{\mathrm{min}}|$]}\quad & \mbox{at $b=0.340$},\\
    \kappa_c &= 0.18845(2)&\quad& \mbox{[PCAC]},&\quad  0.18856(5) \quad\mbox{[|$\lambda_{\mathrm{min}}|$]}\quad & \mbox{at $b=0.350$},\\
    \kappa_c &= 0.18417(3)&\quad& \mbox{[PCAC]},&\quad  0.18417(4) \quad\mbox{[|$\lambda_{\mathrm{min}}|$]}\quad & \mbox{at $b=0.360$}.
\end{alignat}
Although we use the PCAC mass at the largest $N$ for each $b$ and
the limit of $N\to \infty$ is not taken, the critical $\kappa_c$ determined
from both methods are consistent.

\section{Summary}
\label{sec:3}
We have generated ensembles for the reduced matrix model with one adjoint Majorana fermion and twisted gauge boundary conditions, with the 
aim of studying the large $N$ limit of $\mathcal{N} = 1$ SUSY SU($N$) Yang-Mills theory.
We simulated three different bare couplings at $b=0.360, 0.350, 0.340$ and several values of $N$. 
We have analyzed the point of SUSY restoration by looking at the spectrum of $Q_W$ and also at the PCAC mass in the non-singlet 
flavor pseudo-scalar--axial vector channel.
We have found consistency in the value of the critical hopping parameter obtained  from these two determinations.

\section*{Acknowledgments}
P.B., M.G.P. and A.G.-A. acknowledge financial support from the MINECO/FEDER grant PGC2018-094857-B-I00 and
the MINECO Centro de Excelencia Severo Ochoa Program SEV-2016-0597.
This publication is supported by the European project H2020-MSCAITN- 2018-813942 (EuroPLEx) and
the EU Horizon 2020 research and innovation programme, STRONG- 2020 project, under grant agreement No 824093.

M.O. is supported by JSPS KAKENHI Grant Numbers 21K03576 and 17K05417.
K.-I.I. is supported by MEXT as ``Program for Promoting Researches on the Supercomputer Fugaku''
(Simulation for basic science: from fundamental laws of particles to creation of nuclei, JPMXP1020200105) and JICFuS.
This work used computational resources of SX-ACE (Osaka U.) and Oakbridge-CX (U. of Tokyo)
through the HPCI System Research Project (Project ID: hp210027, hp200027, hp190004).

\end{document}